\documentclass[aps,prb,twocolumn,superscriptaddress,showkeys,longbibliography,nobibnotes]{revtex4-2}
\pdfoutput=1 
\usepackage{textcomp,gensymb}
\usepackage{graphicx}
\usepackage[T1]{fontenc}
\usepackage{txfonts}
\usepackage[colorlinks=true,linkcolor=blue,urlcolor=blue,citecolor=blue,anchorcolor=blue]{hyperref}

\begin{document}
	
\title{Disorder driven cluster glass state in a geometrically frustrated hexagonal perovskite}

\author{Shruti Chakravarty}
\affiliation{Department of Physics, Indian Institute of Science Education and Research, Pune, India}

\author{\O ystein Slagtern Fjellv\aa g}
\affiliation{Laboratory for Neutron Scattering and Imaging, Paul Scherrer Institut, Villigen CH-5232, Switzerland}
\affiliation{Department for Hydrogen Technology, Institute for Energy Technology, NO-2027 Kjeller, Norway}

\author{Arpan Bhattacharyya}
\affiliation{Saha Institute of Nuclear Physics, 1/AF Bidhannagar, Kolkata, India}
\altaffiliation[Current address: ]{S N Bose National Center For Basic Sciences, JD Block, Salt Lake Sector 3, Kolkata 700106}

\author{Lukas Keller}
\affiliation{Laboratory for Neutron Scattering and Imaging, Paul Scherrer Institut, Villigen CH-5232, Switzerland}

\author{Sunil Nair}
\email[Correspondance e-mail: ]{sunil@iiserpune.ac.in}
\affiliation{Department of Physics, Indian Institute of Science Education and Research, Pune, India}

\date{\today}

\begin{abstract}
We report the observation of cluster glass-like properties in a double perovskite ruthenate Ba$_2$CoRuO$_6$ through structural (neutron and synchrotron X-ray diffraction), magnetic and transport measurements. The system exhibits classic glassy characteristics like a frequency dependence in ac susceptibility, aging and memory effects along with persistance of short-range correlations upto room temperature. The significant ($\sim30\%$) anti-site disorder on the dimer sites, coupled with the inherent geometrical frustration, allows a variety of exchange (both antiferro- and ferromagnetic) interactions to be distributed randomly across the lattice. On cooling, locally dominant interactions cause spins to nucleate and form local, short-range ordered clusters which grow in size until a global freezing occurs at about $T_f \sim 43K$. 
\end{abstract}

\pacs{}

\maketitle

\section{Introduction}

Double perovskite oxides of the type A$_2$BB'O$_6$ have been extensively studied as a fertile arena to probe emergent magnetic and electric phenomena for many years now. Ideally, a double perovskite is generated by substituting exactly half of the B-sites with a different cation resulting in a three dimensional rock-salt order. However, the matching between ionic radii of the A- and B-cations is crucial towards achieving this perfect order and is parameterized by the Goldschmidt tolerance factor, $t=\frac{r_A+r_O}{\sqrt{2}(\frac{r_B+r_{B'}}{2}+r_O)}$, where $r_A$, $r_B$, $r_{B'}$ and $r_O$ are ionic radii of A, B, B' and O, respectively. When $t<1$, tilting of the BO$_6$ octahedra occurs to compensate for the cation mismatch. When $t>1$ (like for Ba$_2$CoRuO$_6$, $t = 1.071$), the strain is too large, tilting is not enough and face-sharing becomes more common \cite{Vasala2015}. The 6H perovskite structure is most commonly encountered when studying triple perovskite layered systems with the chemical formula A$_3$BB'$_2$O$_9$ where the B and B' cations have two crystallographically distinct positions. They contain face-sharing octahedra or dimers where strong orbital overlap between the B' cations can occur and compete with the intra- and inter-octahedra superexchange interactions. In some 6H perovskites, due to the strong hybridization between the B' and O orbitals, the B'$_2$O$_9$ dimers are suggested as the relevant ``molecular units'' instead \cite{Kimber2012,shlyk2007}. This debate between whether face-shared dimers can be regarded as molecules with defined molecular orbitals or not, is an inherently fascinating feature of 6H Perovskites \cite{Nguyen2020}.\\ 
The $ 3d/4d $ combination of B and B' cations is particularly interesting due to the difference in the spatial extent of the $ 3d $ and $ 4d $ orbitals as well as the increased importance of spin orbit coupling in the latter. Ru$^{5+}$ is a particularly intriguing $ 4d $ ion since it often treads the line between localized and delocalized electron behaviour \cite{Vasala2015} due to both the high correlation energy of the $t_{2g}^3$ configuration and a wider $ 4d $ valence band. The freedom to tune the dimers from a localized (superexchange) to the itinerant (direct exchange or metal-metal bonding) limit enables Ru-based perovskites to display a wide array of interesting electronic and magnetic phenomena including orbital/charge order \cite{Kimber2012}, frustrated, molecular quantum magnets \cite{ziat2017}, large zero-field splitting \cite{chen2019}, and orbital selective Mott insulating behaviour at the molecular level \cite{chen2020}. Incommensurate magnetic structures \cite{Aczel2014}, glassy dynamics and magnetic frustration \cite{Xia2008,Phatak2013,Carlo2013}, colossal dielectric constants \cite{Yoshii2006}, spin-liquid behaviour \cite{Terasaki2017} and even superconductivity \cite{DeMarco2000} has been observed in Ru-based double perovskite systems. 
  
Ba$_2$CoRuO$_6$ mimics this 6H structure with a random distribution of Co and Ru within the face-sharing octahedra, and can be written as Ba$_3$Co[Co$_{0.25}$Ru$_{0.75}$]$_2$O$_9$. Thus, it is evident that Ba$_2$CoRuO$_6$ inherently possesses significant site disorder, which gives rise to complex magnetic behaviour at low temperature. In an attempt to fully interpret this influence of disorder on the magnetic ground state, we have synthesized this material and extensively characterized it through structural, magnetic, dielectric, thermal \& electronic transport, along with temperature dependent neutron \& synchrotron measurements.  

\section{Experimental Methods}

\begin{figure}[ht]
	\includegraphics[trim=1cm 1cm 1cm 1cm, width=\columnwidth]{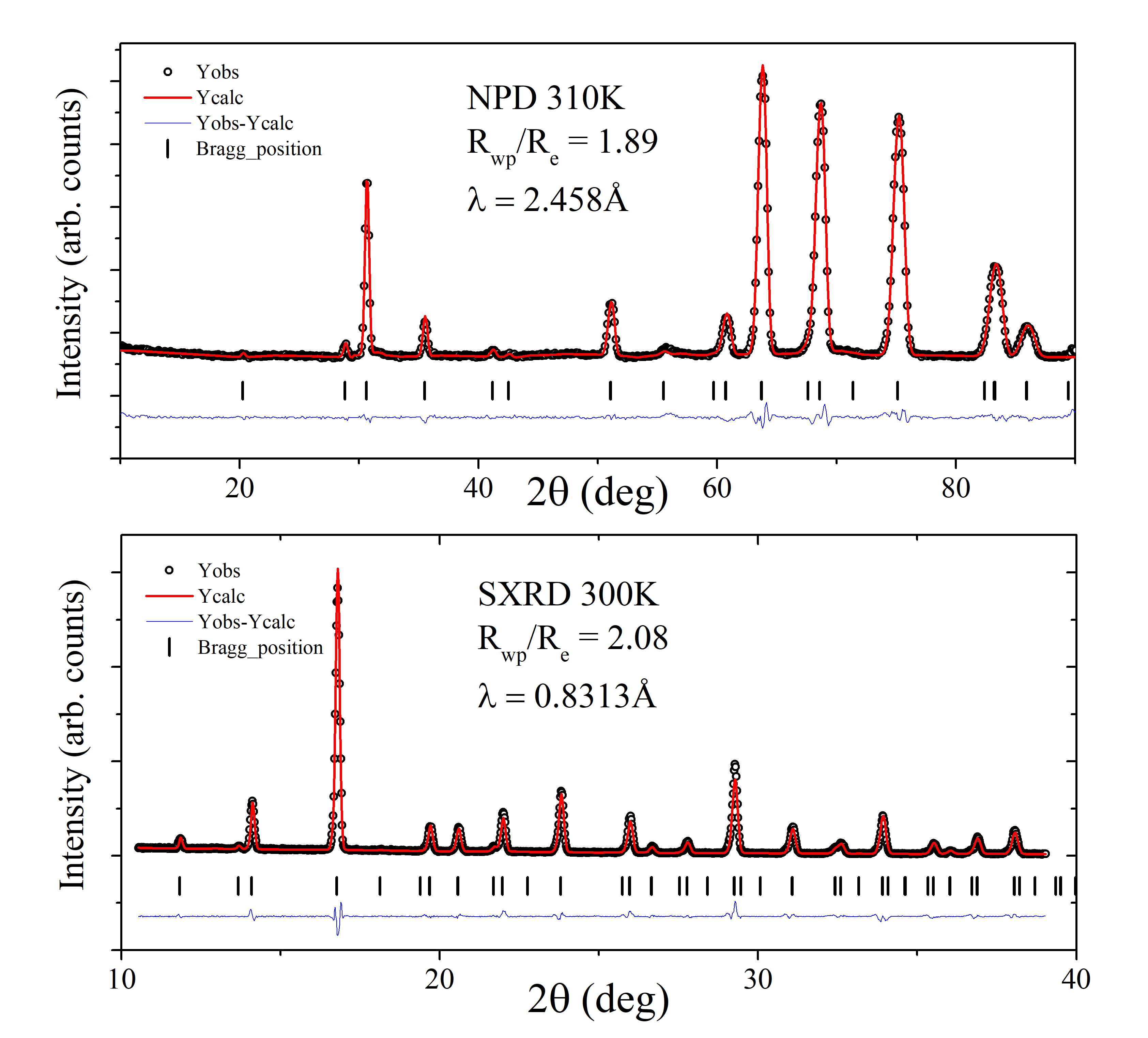}
	\caption{Neutron Powder Diffraction (NPD) (top) and Synchrotron X-ray Diffraction (SXRD) (bottom) profiles at room temperature refined using FULLPROF Suite.}
	\label{diffraction} 
\end{figure}

Polycrystalline samples of Ba$_2$CoRuO$_6$ were prepared via solid state synthesis method using stoichiometric amounts of raw materials (BaCO$_{3}$, Co$_{3}$O$_{4}$ and RuO$_{2}$) following the procedure detailed in \cite{kim1995}. The raw materials were mixed and ground in a mortar using ethanol until a fine, homogeneous mixture was obtained. The mixture was then heated at 1150\degree C twice, for 24 hours each with intermediate regrinding and pelletization. Laboratory X-ray Diffraction (XRD) measurements at room temperature performed using Bruker D8 Advance Diffractometer (CuK$_{\alpha}$ $\lambda$=1.5406\AA) confirmed the sample to be pure and single-phase.
Elemental compositions and their homogeneity were reconfirmed by using an energy dispersive X-ray spectrometer (Ziess Ultra Plus). The average stoichiometry calculated from statistical analysis of the EDS data is Ba$_{2}$Co$_{0.97}$Ru$_{1.03}$O$_{6}$ within error limits, affirming that our sample is homogeneous and single phase. Temperature dependent Synchrotron XRD (SXRD) measurements were performed at the Indian Beamline (PF-18B) at Photon Factory, KEK, Japan using an X Ray wavelength of 0.8313\AA{}. Neutron Powder Diffraction (NPD) was performed at DMC (Cold Neutron Diffractometer) at Paul Scherrer Institute (PSI) using a neutron beam wavelength of 2.458\AA{}. Rietveld and Le-Bail refinements were performed using the \textit{FULLPROF SUITE} \cite{fullprof} as well as \textit{JANA2020} \cite{jana} to analyse the diffraction data and obtain structural information. Crystal Structures were generated using VESTA \cite{vesta}.
Specific Heat and Magnetic measurements were made using Quantum Design PPMS and MPMS-XL SQUID Magnetometer, respectively. Resistivity and Dielectric measurements were performed on a homemade setup. Dielectric spectroscopy was performed using Novocontrol Alpha-A Impedance Analyzer in the frequency range of 1Hz-1MHz.

\section{Results and Discussion}

\begin{figure}[ht]
	\includegraphics[width=\columnwidth]{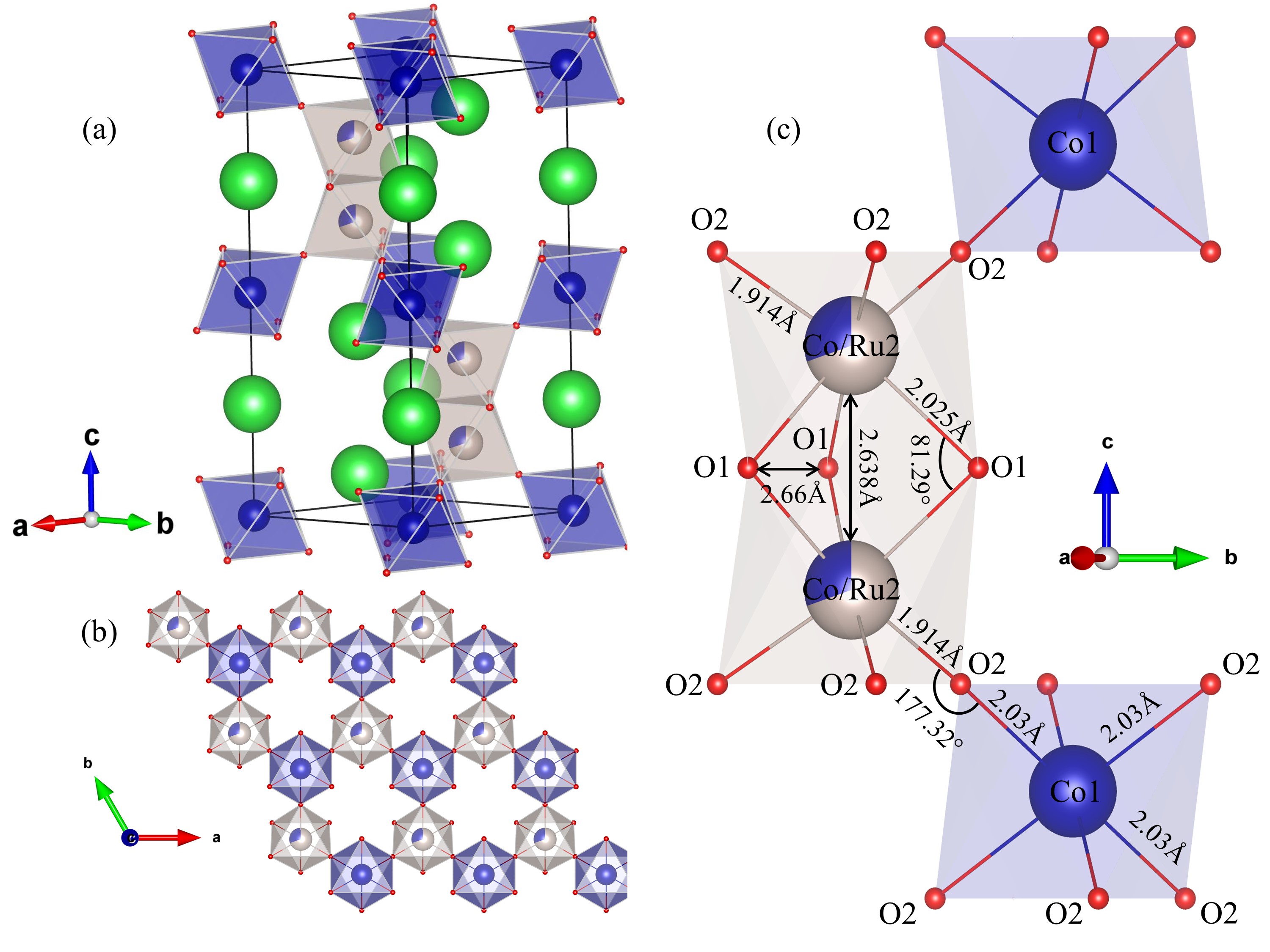}
	\caption{(a) Hexagonal structure of Ba$_{2}$CoRuO$_{6}$ showing corner and face-shared octahedra, (b) Buckled honeycomb lattice formed by the corner sharing of B and B' ions, (c) Bond lengths and bond angles in the (Co/Ru)$_{2}$O$_{9}$ dimers and CoO$_{6}$ octahedra}
	\label{structure}
\end{figure}

All peaks of the SXRD and NPD patterns of Ba$_2$CoRuO$_6$ were indexed and refined with the 6H structure in space group $P6_3/mmc$ (Fig. \ref{structure}), with lattice parameters $a=b=5.700(4)$\AA{} and $c=13.970(1)$\AA{} consistent with previous reports \cite{kim1995,Hira2021}. Structural parameters calculated at room temperature from synchrotron and neutron diffraction data were found to be consistent confirming the quality of refinement (Fig.\ref{diffraction}) and are tabulated in Table \ref{strucpar}. Oxygen positions were refined only using neutron data and no significant variation in oxygen occupancies was observed. We find that Co$^{3+}$ fully occupies the corner sites and forms layers of triangular lattices perpendicular to the c-axis (Fig.\ref{structure}(b)) while there is significant anti-site disorder in the dimer $4f(\frac{1}{3},\frac{2}{3},z)$ sites with Co$^{3+}$ occupying about 30\% of them. These sites form their own triangular layer in the a-b plane and the two corner-sharing octahedra also form a buckled honeycomb lattice. 

\begin{table}
	Space Group: P6$_{3}$/mmc (Hexagonal) \\
	a = b = 5.700(4)\AA{} , c = 13.970(1)\AA{}, $\alpha$ = $\beta$ = 90\degree , $\gamma$ = 120\degree \\
	
	\begin{ruledtabular}
		\begin{tabular}{ccccccc}
			\textbf{Atom} & \textbf{Wyckoff} & \textbf{x} & \textbf{y} & \textbf{z} & \textbf{B} & \textbf{Occ}  \\
			\hline
			Ba1 & 2b & 0 & 0 & 1/4 & 0.0005 & 1 \\
			Ba2 & 4f & 1/3 & 2/3 & 0.90963 & 0.0005 & 1 \\
			Co1 & 2a & 0 & 0 & 0 & 0.24289 & 1 \\
			Co2 & 4f & 1/3 & 2/3 & 0.15561 & 0.21454 & 0.30\\
			Ru2 & 4f & 1/3 & 2/3 & 0.15561 & 0.21454 & 0.70\\
			O1 & 6h & 0.48892 & -0.02223 & 0.25 & 0.25418 & 0.97\\
			O2 & 12k & 0.16892 & 0.33811 & 0.41713 & 0.52233 & 1\\			
		\end{tabular}
	\end{ruledtabular}
	
	\caption{\label{strucpar} Structural Parameters of Ba$_{2}$CoRuO$_{6}$ obtained from Rietveld refinement of the neutron diffraction pattern measured at 310K}
\end{table}

Table \ref{bonds} lists all the bond lengths and angles calculated from the refined structure and  Fig. \ref{structure}(c) visualizes them. The average distance between the dimer cations obtained from the refinement are 2.638\AA{} which is smaller than that Ru\textendash Ru metallic bond length (2.6725\AA{} \cite{Jain2013}), implying that the intradimer antiferromagnetic (AFM) direct exchange have significant strength. The average (Ru/Co2)\textendash O distance is $\sim1.97$\AA which matches almost perfectly with the expected value for a Ru$^{5+}$\textendash O$^{2-}$ bond \cite{Gagne2020}, thus confirming that the material has a Co$^{3+}$/Ru$^{5+}$ distribution instead of Co$^{4+}$/Ru$^{4+}$ or Co$^{2+}$/Ru$^{6+}$. The Ru$^{5+}$ ions are also slightly displaced from their centers due to Coulomb repulsion as is evident from the difference in the ((Ru/Co2)\textendash O1 $\sim 2.0246$\AA) and ((Ru/Co2)\textendash O2 $\sim1.9141$\AA) bond lengths.

The spin-state flexibility of Co$^{3+}$ opens up the possible existence of low-spin (LS), high-spin (HS) or intermediate spin (IS). The factors affecting the spin-state of Co here are the oxidation state, surrounding crystal field, coordination number and the local symmetry and the type of neighbouring ions. A LS Co$^{3+}$ is well documented and has been found in: LiCoO$_2$ \cite{oku1978}, La$_2$Li$_{0.5}$Co$_{0.5}$O$_4$ \cite{hu1998}. HS Co$^{3+}$ has been documented in weaker crystal fields like YBaCo$_4$O$_7$ with a tetrahedral local symmetry \cite{hollmann2009}, and in Sr$_2$CoO$_3$Cl \cite{hu2004} and BiCoO$_3$ \cite{belik2006} with a square pyramidal local symmetry. The HS Co$^{3+}$ in octahedral coordination is much rarer and has been seen only in systems with a LS-HS mix (LaCoO$_3$ \cite{haverkort2006}, Pr$_{0.5}$Ca$_{0.5}$CoO$_3$ \cite{tsubouchi2002} ) or with oxygen deficiency (RBaCo$_2$O$_{5.5}$ (R=Rare-earth metal) \cite{martin1997, hu2012}, Sr$_{1-x}$Y$_x$CoO$_{3-\delta}$) \cite{kobayashi2005}. There is also a continuing debate on whether the observed spin state is actually a mixture of HS and LS, or is a pure IS state.

A pressure dependent study of the spin state of the octahedrally coordinated Co$^{3+}$ in the 3d/4d transition metal (TM) oxide SrCo$_{0.5}$Ru$_{0.5}$O$_{2.96}$ \cite{chen2014} concludes that at room temperature and ambient pressure, the Co ions are in a HS state. Applying pressure induces a complete HS $\rightarrow$ LS transformation with no intermediate IS state. The crossover lies at about 1.93\AA{} Co-O bond length where the HS and LS states are strongly mixed and local lattice relaxations are allowed. Calculations using the configuration interaction cluster model indicate that the extremely large distortions are needed to stabilize an IS state. Another study on the evolution of the spin state of Co$^{3+}$ in CoO$_6$ octahedral environment in Sr$_2$Co$_{0.5}$Ir$_{0.5}$O$_4$ \cite{chin2017} also finds that for an IS state to be stabilized there should be a strong elongated tetragonal distortion along with a short in-plane Co-O bond length. For in-plane Co-O bonds larger than 1.9\AA{} the IS state is highly unfavourable. A stable HS state is seen at high temperatures and ambient pressures (Co-O in plane = 1.944\AA, Co-O apex = 2.025\AA) which transforms completely into an LS state (Co-O in plane = 1.815\AA, Co-O apex = 1.925\AA) beyond 9.7GPa again without encountering the IS state at all. In LaCoO$_3$, the Co-O bond length decreases from 1.9329\AA{} to 1.888\AA{} with increasing pressure, crossing a mixed HS-LS state to a pure LS state. The mixed spin state is due to shorter bond lengths (Co-O in plane = 1.967\AA, Co-O apex = 2.02\AA), close to the boundary of HS and LS. Clearly, the correlation between the spin state of Co and its distance from O is an effective indirect way to gauge the spin state of Co$^{3+}$ ions.

In this context, Ba$_2$CoRuO$_6$ is most likely to stabilize a HS state since there is no distortion in either of the two distinct octahedra encountered here with fairly large average bond lengths. The corner-shared CoO$_6$ octahedra contain equidistant in- and out of- plane Co-O bonds (2.0315\AA). The octahedra that is shared also has equal in- and out of-plane average bond lengths (1.9694\AA). These values are close to the HS state bond distances in Sr$_2$Co$_{0.5}$Ir$_{0.5}$O$_4$ and SrCo$_{0.5}$Ru$_{0.5}$O$_{2.96}$, implying that here too, the crystal field is fairly weak and is likely to favour an HS state at high temperatures. The stoichiometry calculated from the neutron diffraction analysis is Ba$_2$Co$_{1.03}$Ru$_{0.97}$O$_{6-\delta}$ with $\delta=0.07$ which points to a slight oxygen deficiency similar to SrCo$_{0.5}$Ru$_{0.5}$O$_{2.96}$ and validating our claim of an expected HS state for Co$^{3+}$ in this compound. Cooling the sample causes only marginal changes in the bond lengths indicating no significant change in Co$^{3+}$ spin states on cooling and the octahedra remain regular with the average in-plane and apex bond lengths Co1 octahedra = 2.0226\AA\ and in Co/Ru2 octahedra = 1.9617\AA. Thus, we can conclude that the Co ions remain in a HS state between 2-300K in this system.

\begin{table}
	\begin{ruledtabular}
		\begin{tabular}{ccc}
			Ba1 - O1  & $\times6$ & 2.8527(3)\AA \\
			Ba1 - O2  & $\times6$ & 2.8703(2)\AA \\
			Ba2 - O1  & $\times3$ & 2.838(3)\AA \\
			Ba2 - O2  & $\times6$ & 2.8525(3)\AA \\
			Ba2 - O2'  & $\times3$ & 2.914(3)\AA \\
			Co1 - O2 & $\times6$ & 2.0315(1)\AA \\
			(Ru/Co2) - O1 & $\times3$ & 2.0246(2)\AA \\
			(Ru/Co2) - O2 & $\times3$ & 1.9141(2)\AA \\
			(Ru/Co2) - O (avg) & & 1.9693(2)\AA \\
			Co1 - (Ru/Co2) & & 3.9445(2)\AA \\		
			(Ru/Co2) - (Ru/Co2) & & 2.638(6)\AA \\
			\\
			O1 - (Ru/Co2) - O1 & & 82.13(9)\degree \\
			O1 - (Ru/Co2) - O2 & & 91.39(1)\degree \\
			O2 - (Ru/Co2) - O1 & & 171.41(1)\degree \\
			O2 - Co1 - O2 & & 90.74(4)\degree \\
			Co1 - O2 - (Ru/Co2) & & 177.32(8)\degree \\
			(Ru/Co2) - O1 - (Ru/Co2) & & 81.29(1)\degree \\
		\end{tabular}
	\end{ruledtabular}
	
	\caption{\label{bonds}Bond Angles and Lengths for Ba$_{2}$CoRuO$_{6}$ obtained from Rietveld refinement of neutron diffraction data collected at 310K}
\end{table}

The Co/Ru2\textendash O\textendash Co/Ru2 superexchange $(\angle_{ROR}=81.29\degree)$ competes with the intradimer direct exchange by favoring a weak ferromagnetic (FM) coupling. Direct exchange between Co$^{3+}$ ions is not expected since the $3d$ orbitals are not extended enough. For the interactions between (Ru$^{5+}$:$t_{2g}^3$) and (Co$^{3+}$:$t_{2g}^4e_{g}^2$), depending on which orbital of Co participates in exchange ($t_{2g}/e_{g}$), the interaction maybe FM/AFM, respectively. There will also be finite contribution from next-nearest neighbour (NNN) exchange. All of these competing interactions are randomly distributed throughout the lattice due to the antisite disorder, implying that exchange randomness is a key player in dictating the physics of this system. The O1\textendash O1 distance is small $\sim 2.661$\AA and allows the shared face of the octahedra to effectively shield the cationic repulsion \cite{Santoro2000}, similar to other 6H Perovskites \cite{Lightfoot1990,Kayser2017}. 

\begin{figure}[ht]
	\includegraphics[width=\columnwidth]{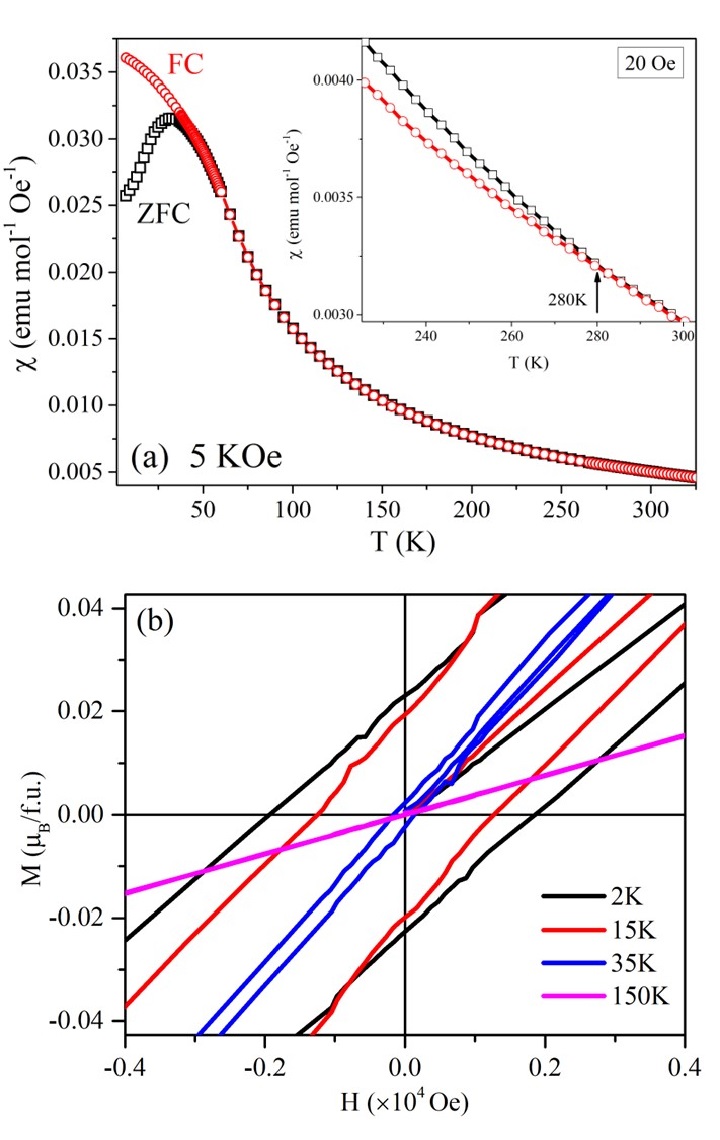}
	\caption{\textit{Top}: Temperature dependent dc Susceptibility ($\chi_{dc}$) measured at 5KOe showing bifurcation between ZFC and FC curves below 47K, Inset: $\chi_{dc}$ measured at $H = 20$ Oe showing the separation of ZFC and FC curves at 280K, \textit{Bottom}: $M(H)$ curves showing small, but finite hysteresis below $T_f$ indicating a weak FM contribution.}
	\label{MT}
\end{figure}
\begin{figure}[ht]
	\includegraphics[trim=1cm 1cm 0 1cm, width=\columnwidth]{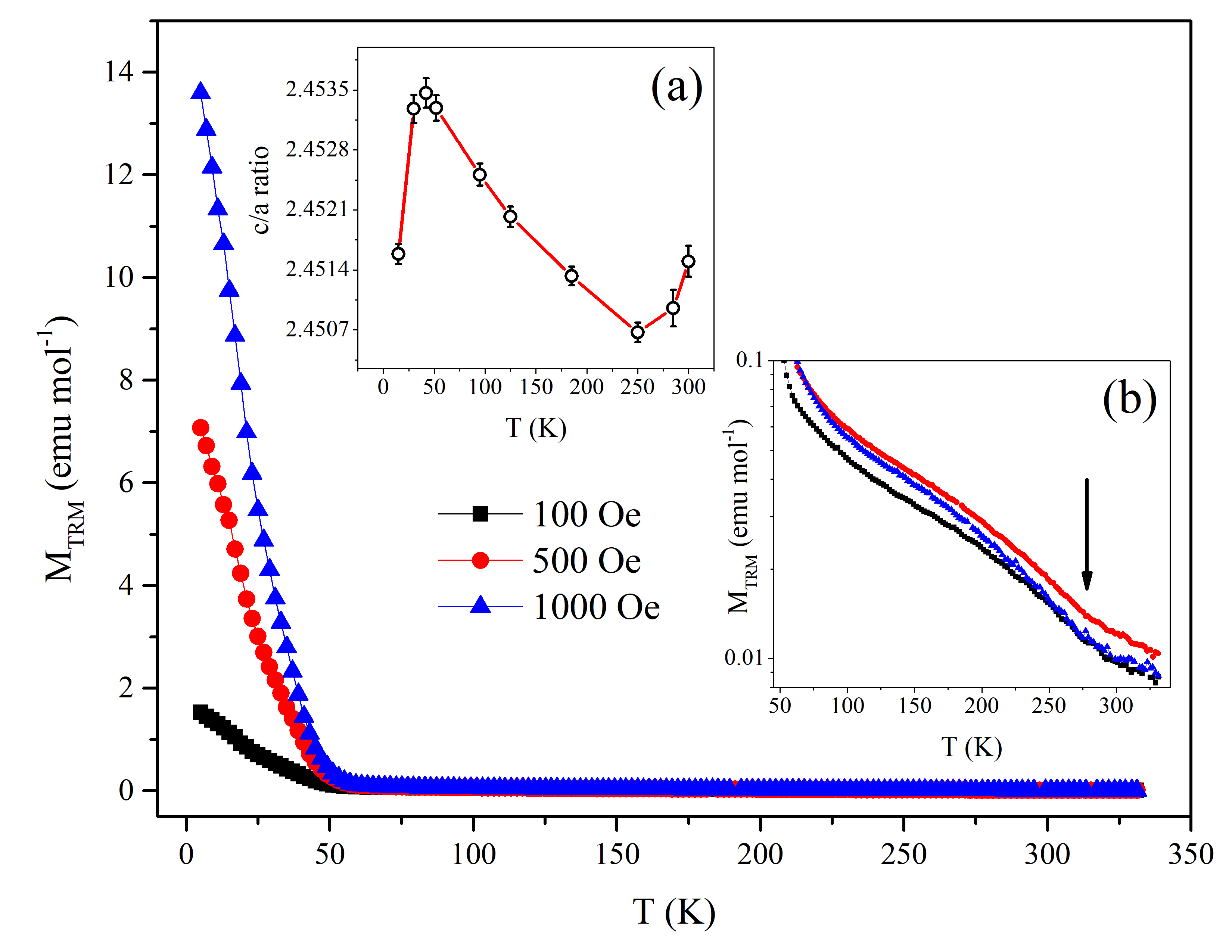}
	\caption{Thermo-Remanent Magnetization ($M_{TRM}$) vs Temperature (T) measured in zero field in warming mode (using three different cooling fields), (a) Thermal variation in lattice parameters extracted from SXRD data showing features at 47K and 250K, (b) semilog plot of M$_{TRM}$ vs T highlighting the subtle change of slope at 280K (indicated using the arrow).}
	\label{highT}
\end{figure}


dc magnetic susceptibility ($\chi_{dc}$) of the sample measured at 5KOe (Fig.\ref{MT}), shows a cusp in the Zero-Field Cooled (ZFC) curve accompanied by a bifurcation between the ZFC and Field Cooled (FC) curves around $T_{f}=43K$ reminiscent of a spin-glass (SG) like behaviour \cite{binder1986}. Canonical spin glasses generally show a temperature independent FC magnetization ($M _{FC} $) below $T_{f}$ and this feature is often used to distinguish them from a superparamagnet (SPM) \cite{Giot2008}. However, it is also possible for SPMs with a narrow volume distribution to show this plateau below the blocking temperature \cite{Nair2007}. On the contrary, our FC curve continues to increase below $T_f$, a behaviour also documented in various cluster glasses (CGs) \cite{Pejakovic2000,Koyano2013,Mukherjee1996}. Attempts to fit $\chi_{dc}^{-1}(T)$ to the Curie-Weiss (CW) law at high temperature failed (not shown here), as we encountered varying values of the Weiss temperature ($\theta_{CW}$) and Curie constant ($ C $) on changing the fitting (temperature) range, indicative of a rounding of $\chi^{-1}$. This deviation from the Curie-Weiss (CW) linearity could be suggestive of clustering \cite{Giot2008}, implying that the system is not fully paramagnetic at room temperature. $\chi_{dc}$ at low fields ($H=20$ Oe) shows that irreversibility occurs at temperatures as high at 280K (inset (a) of Fig.\ref{MT}). In both SGs and SPMs, existence of a finite dipolar interaction can cause the ZFC and FC curves to split at temperatures much higher than $T_{F}$ (or $T_{B}$) and continue to rise as the sample is cooled \cite{Nair2007,Luo1991}. This separation of ZFC and FC curves above the ZFC maxima with growing $M_{FC}$ below the bifurcation point has also been observed in many other cluster glasses \cite{Middey2011,Nair2007,Freitas2001,Pejakovic2000,Koyano2013,Shand2012}. The temperature variation of lattice parameters (inset (a) of Fig.\ref{highT}) obtained from Le-Bail fits of synchrotron patterns measured at different temperatures also shows sharp changes at 50K and 280K consistent with the magnetization data, the former reflecting the freezing transition and the latter associated with the onset of magnetic correlations. We pick up a signature of these short-range correlations at $\sim 280K$ in the Thermo-Remanent Magnetization (TRM) measurements (Fig.\ref{highT}) as well, which have been useful in studying glassy behaviour \cite{Mathieu2001,Kumar2017,Garg2022}. In this experiment, the sample is cooled to $T<T_f$ in presence of an external field and then the field is turned off. $M_{TRM}(T)$ is then measured keeping $H = 0$ in warming mode. TRM being a zero-field measurement allows us to observe the thermal evolution of the magnetic state while being more sensitive to the antiferromagnetic order within clusters which would otherwise be suppressed by the paramagnetic contribution. Along with the huge upturn at $T_f$, we clearly capture a small change of slope around 280K  (inset (b) of Fig.\ref{highT}) confirming that nucleation of spin clusters begins close to room temperature. The field dependence of $M_{TRM}$ in this region is unlike that below $T_f$ due to the fact that being a zero field measurement, at such low moment values, the effect of the remnant field of the superconducting magnet of the MPMS-SQUID becomes relevant. Careful high temperature magnetic susceptibiity measurements would be essential to access the true paramagnetic region for this system.

\begin{figure}[ht]
	\includegraphics[trim=2cm 1cm 1cm 0, width=\columnwidth]{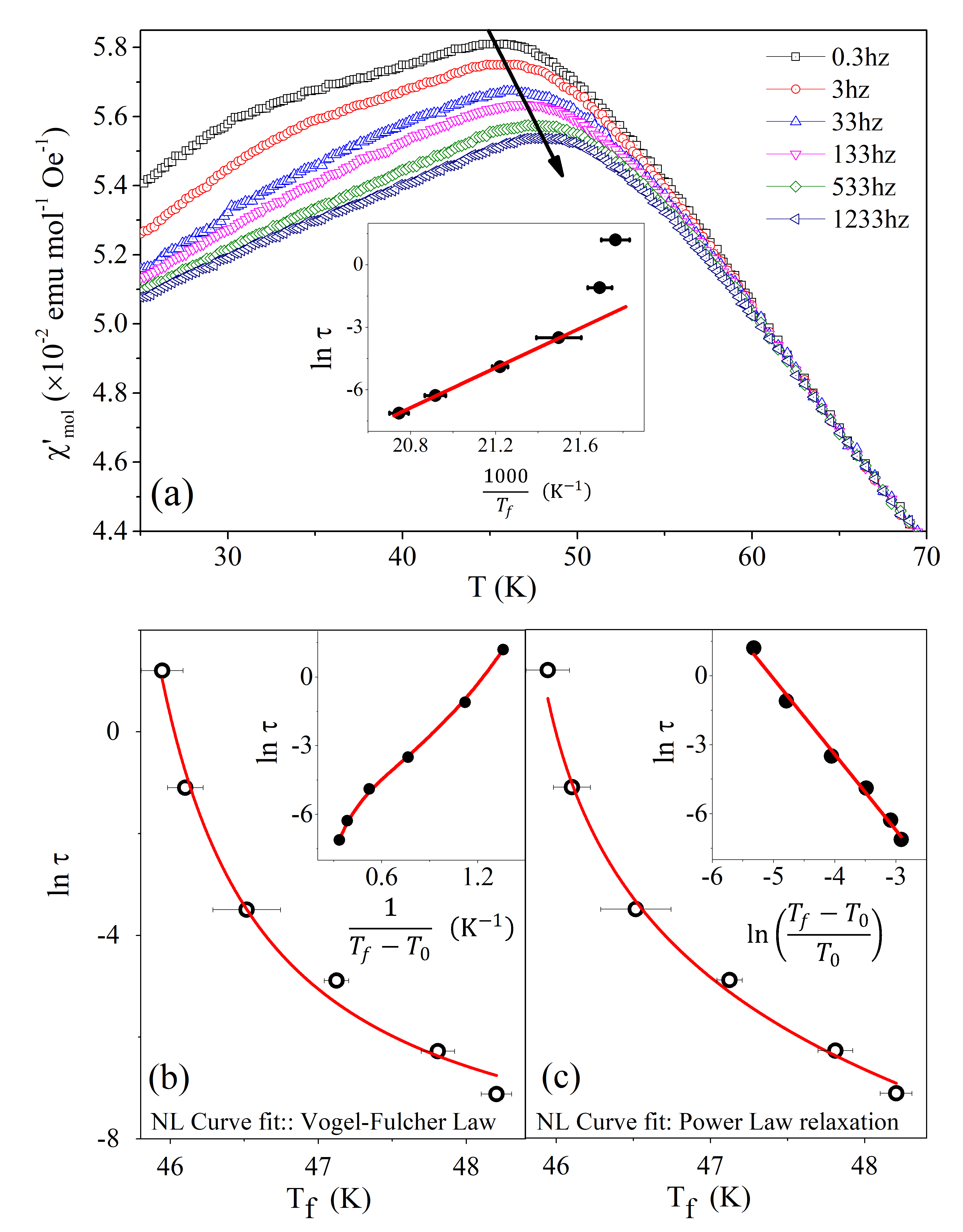}
	\caption{(a) Real part of $\chi_{ac}$ as a function of temperature for different frequencies. Inset: ln$\tau$ plotted against $1000/T_{f}$ showing non-linear (non-Arrhenius) behaviour, (b) ln$\tau$ vs $T_f$ nonlinear curve fit to the Vogel-Fulcher Law, Inset: ln$\tau$ 1/($T_f-T_0$) using $T_0$ obtained from nonlinear fit showing non-linearity, (c) Power law fit to  ln$\tau$ vs $T_f$, Inset: the linear scaling of power law fit.}
	\label{ACchi} 
\end{figure}

To confirm if the ZFC-FC bifurcation in the dc susceptibility indeed corresponds to a cooperative freezing of spins, we measure the frequency dependence of the ac magnetic susceptibility ($\chi_{ac}$) with temperature. Fig.\ref{ACchi}(a) shows the real part of $\chi_{ac}$ measured with an excitation field of $H=2.6$ Oe after cooling the sample in zero field. The temperature of the ZFC magnetization ($M_{ZFC}$) peak shows a small shift with frequency attesting to the glassiness of this transition. This shift is quantified in the Mydosh Parameter ($\delta T_f$),
\begin{equation}
	 \delta T_f = \frac{\Delta T_f}{T_f \Delta (log \omega)}
\end{equation}
which in our case is 0.0136. Here, $T_f$ is the temperature at which the cusp occurs in Re($\chi_{ac}$) at a certain angular frequency ($\omega = 2 \pi f$, $f$ being the applied frequency in hz). This value is higher than typical canonical spin glasses like CuMn (0.005) but matches well with values observed for reentrant spin glasses \cite{Mydosh1993} and cluster glasses \cite{Lago2014,Shand2012,Marcano2007,Harikrishnan2008}. This distinguishes it from a SPM where this frequency shift is expected to be larger and $\delta T_f$ is $>0.1$ \cite{Mydosh1993}. The logarithmic relaxation time ln$\tau$ $(\tau=\frac{1}{\omega})$ also deviates strongly from linearity when plotted against $1000/T_f$ (inset of Fig.\ref{ACchi}(a)) ruling out the possibility of a SPM-like thermally-activated blocking (Arrhenius-like behaviour) of spin clusters. Fitting  ln$\tau$ vs $T_f$ to the Vogel-Fulcher (VF) law, 
\begin{equation}
	\tau=\tau_{0}\cdot exp\bigg({\frac{E_a}{k_{B}(T-T_0)}}\bigg)
	\label{vft}
\end{equation}
gives a better fit (Fig.\ref{ACchi}(b)) resulting in parameters $\tau_0 \approx 10^{-4}$s, $T_0 = 45.21\pm0.11$ K and $E_a = 0.65\pm0.18$ meV. The VF law is an empirical law which is generally used to describe the viscosity of supercooled liquids and real glasses. It is essentially a modified version of the Arrhenius Law, with $T_0$ as an extra parameter. A concrete physical interpretation of $T_0$ however, remains elusive \cite{Mydosh1993}. A linear scaling of the non-linear fit to eq.\ref{vft} reveals the fit to be unsatisfactory (inset of Fig.\ref{ACchi}(b)) and hence, we look to the standard theory of phase transitions where dynamical scaling near the transition expects $\tau$ to vary as:
\begin{equation}
	\tau = \tau_{0} \cdot \bigg(\frac{T_{f}}{T_{0}}-1\bigg)^{-z\nu}
	\label{taueq}
\end{equation}
where, $\tau_{0}$ is the characteristic relaxation time, $T_0$ the transition temperature and $z\nu$ the dynamical exponent that encapsulates the critical slowing down of relaxation dynamics near the phase transition.
A non-linear fit to eq.\ref{taueq} and its linear scaling is shown in Fig.\ref{ACchi}(c) and its inset respectively. It is evident that the linear fit is drastically improved for eq.\ref{taueq}. The parameters obtained from the fit are: $T_0 = 45.72 \pm 0.03$ K, $\tau_0 = 6.22 (\pm 0.19) \times 10^{-8} $s and $z\nu = 3.29 \pm 0.12 $. When compared to typical values for a canonical spin glass CuMn (4.6 at.\%) ($\tau_0 = 10^{-12} s$ and $z\nu = 5.5 $) we can see that the relaxation times of the two systems differ by almost 4 orders of magnitude demonstrating that the entities that respond to the external perturbation in this system are not individual spins but correlated, interacting spin clusters. The value of $z\nu$ is also lower than its typically observed range for SGs (4-12) \cite{Mydosh1993}. However, $z\nu<4$ has also been reported in some CG systems \cite{chakrabarty2014,warshi2020,Harikrishnan2008}. A low value of $z\nu$ approaching that for a mean-field phase transition ($z\nu = 2$) has been recognized as a marker of strong interactions allowing cooperative freezing of magnetic moments at $T_f$ \cite{Alonso2010}.

\begin{figure}[ht]
	\includegraphics[trim=1cm 1cm 0 0, width=\columnwidth]{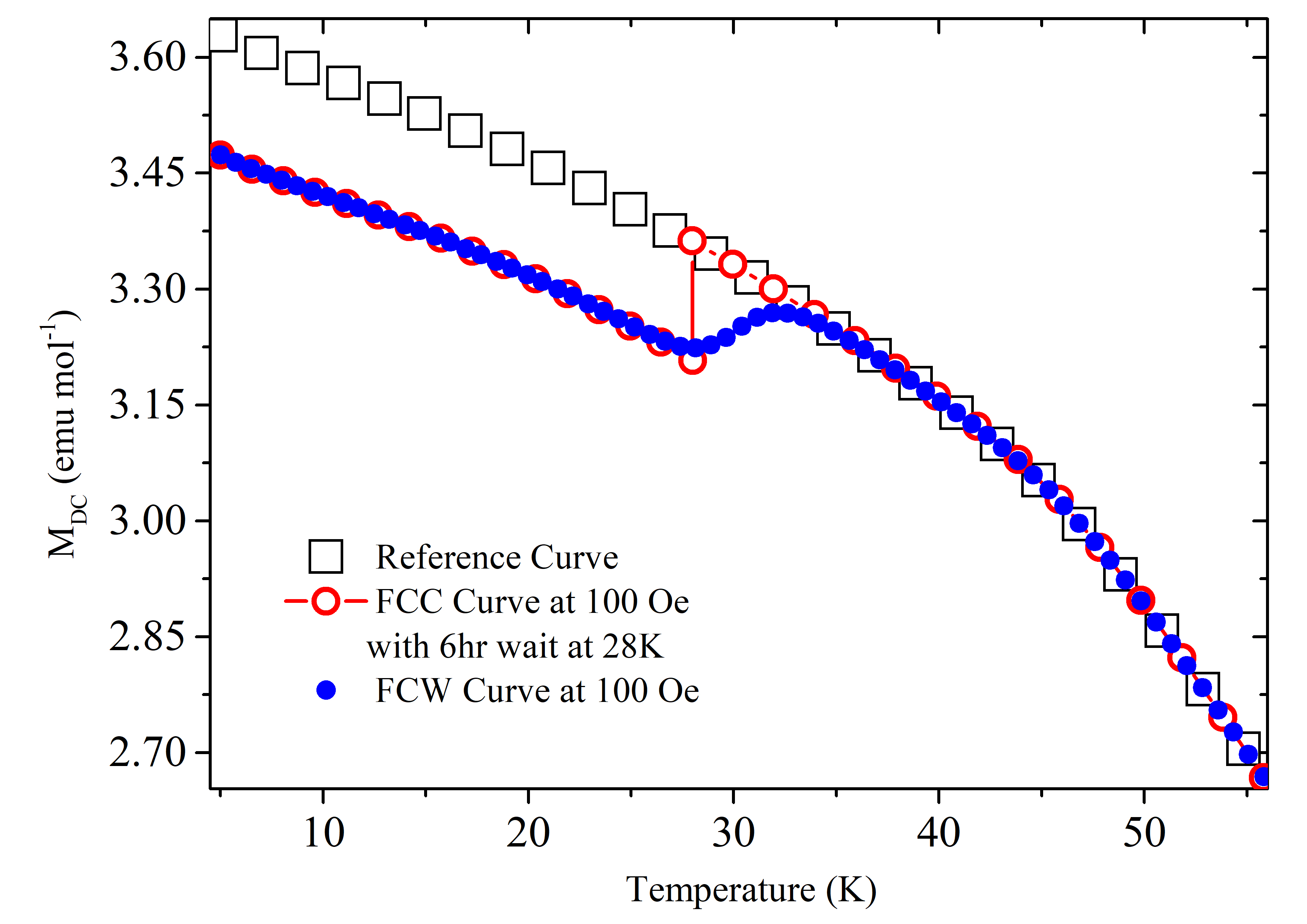}
	\caption{Observation of the memory effect in Ba$ _{2} $CoRuO$ _{6} $ at T = 28K (<$T_{f}$) with a waiting time of 6 hours, using a cooling field of 100 Oe}
	\label{memory} 
\end{figure} 

Ba$_2$CoRuO$_6$ also shows the classic glassy behaviour of retaining the \textit{memory} of an aging process within the frozen state. For this experiment, a reference FC curve is first measured by normally cooling the sample in $H=100$ Oe and then measuring $M_{FC}$ during warming. The sample is then cooled again in field from $T>T_{f}$ but this time, on reaching $T = 28K <T_f$, the field is switched off and the magnetization is allowed to relax. After $t_w =$ 6 hours, the field is switched back on and the sample is allowed to continue cooling normally to the lowest temperature and magnetization is measured during warming. As evident from the inset of Fig.\ref{memory}, the material \textit{remembers} the aging it underwent during cooling and smoothly merges into the previously measured cooling curve revealing (and simultaneously, erasing) the memory \cite{Vincent2007}. This clear evidence of magnetic memory in the system reaffirms the glassiness of the low temperature state of Ba$_2$CoRuO$_6$.

\begin{figure}[ht]
	\includegraphics[trim=1cm 0.5cm 0 0, width=\columnwidth]{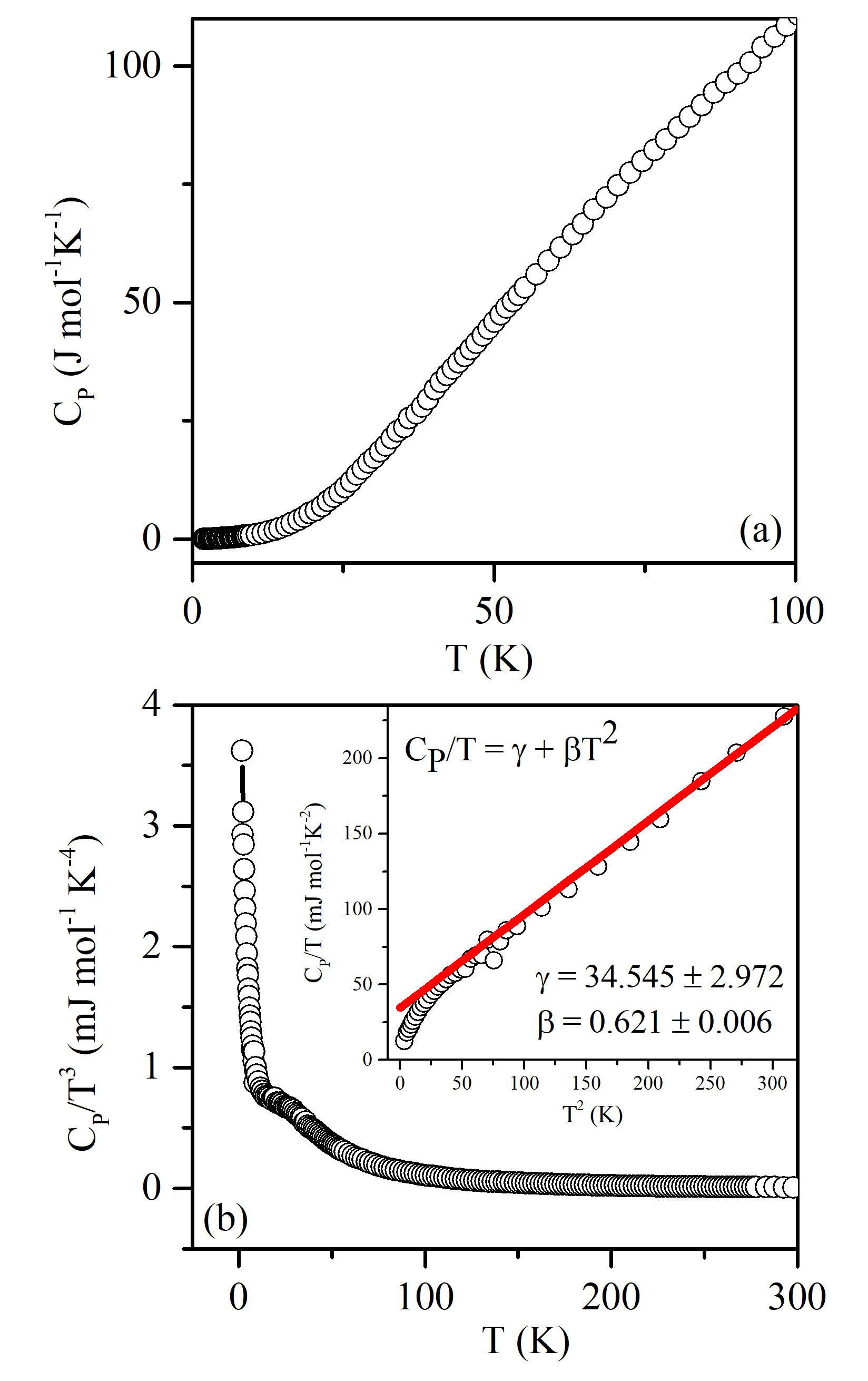}
	\caption{Plots of (a) $C_P$ vs T showing the absence of any feature in the vicinity of $T_f$, and (b) $C_P/T^3$ plotted against $T$ showing a sharp upturn below $T_f$. Inset of (b) shows $C_P/T$ plotted against $T^2$  with its linear fit showing the deviation from the Debye behaviour at low temperatures. The fitting equation as well the obtained fit parameters are also mentioned.}
	\label{spheat} 
\end{figure} 

Further confirmation that the magnetic transition at 43K relates to a global freezing of spin clusters and not long-range ordering, comes from the specific heat of Ba$_2$CoRuO$_6$ measured in zero field, which is featureless around 50K (Fig.\ref{spheat}(a)) implying that the entropy change associated with this transition is negligible. A linear fit to $C_P/T$ vs $T^2$ in the low temperature regime shown in the inset of Fig.\ref{spheat}(b) gives $\gamma = 34.55\pm2.97$ mJ mol$^{-1}$ K$^{-2}$ and $\beta = 0.62\pm0.01$ mJ mol$^{-1}$ K$^{-4}$. The $\gamma$ value is unusually large for an insulating system but this, along with the downturn below $\sim10K$ can also be attributed to the glassy ground state. High values of $\gamma$ have also been reported in various other frustrated systems including 6H triple perovskites \cite{Garg2022,Nag2016,Zhou2011} and half doped manganites \cite{Banerjee2009,Hardy2003}. These parameters allow us to calculate the Debye Temperature ($\theta_D$) using the relation: 
\begin{equation}
	\theta_{D}=\bigg(\frac{12\pi^{4}pR}{5\beta}\bigg)^{1/3}
\end{equation}
where, $p$ is the number of atoms per formula unit (=10 for double perovskites) and $R$ is the Gas Constant. We obtain a value of $\theta_{D}=315.14K\pm1.01$ K, which is close to the typically observed values for perovskites (350-400K) \cite{Hardy2003,Prellier2000}. $C_P/T^3$ plotted against $T$ (Fig.\ref{spheat}(b)) shows an upturn below 10K which is related to a presence of tunneling in two-level systems and is often seen in glassy/phase-separated systems \cite{Garg2022,Garg2021,Banerjee2009}. In Debye materials, where periodicity is expected, $C_P/T^3$ is just expected to flatten out at low temperature. 

\begin{figure}[ht]
	\includegraphics[trim=1cm 1cm 1cm 0, width=\columnwidth]{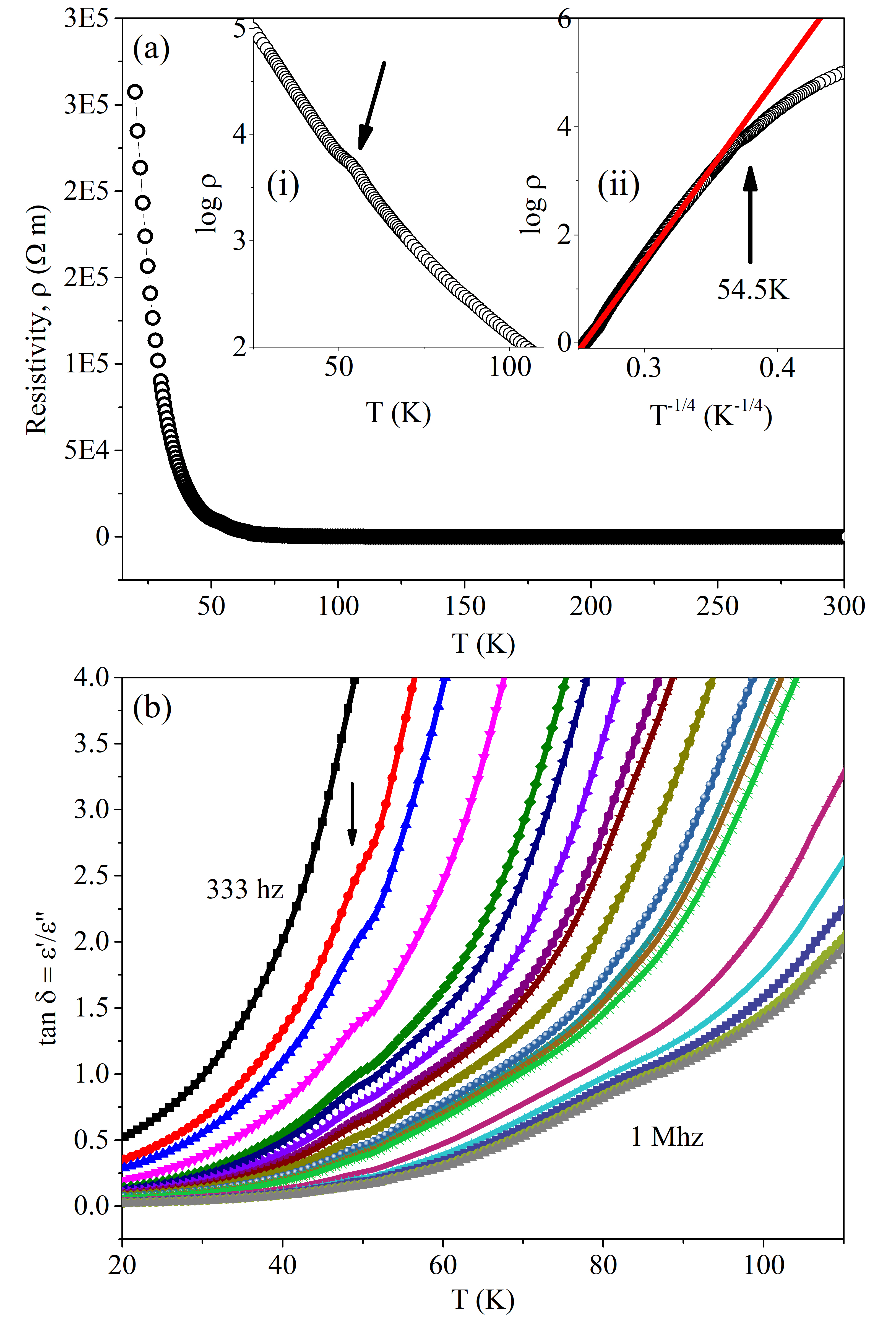}
	\caption{Top: Temperature dependence of resistivity ($\rho(T)$) showing insulating behaviour at low temperatures. Inset: (i) log$\rho$ plotted against T shows a very small feature (indicated by the arrow) near $T_f$, (ii) log$\rho$ plotted against $T^{-1/4}$ showing deviation from linearity below ~55K, implying that the 3D VRH model describes the conductivity well until the glass transition is approached. Bottom: Dielectric loss tangent (tan$\delta=\epsilon'' / \epsilon'$) plotted as a function of Temperature also shows a subtle feature close to 50K hinting at a weak magnetodielectric coupling.}
	\label{rho} 
\end{figure} 

Resistivity measurements show semiconducting behaviour between 20-300K with a small feature in the vicinity of $ T_f $. Linearity upto temperatures approaching $T_f$ is obtained when ln$\rho$ is plotted against $T^{-\frac{1}{4}}$ (inset (ii) of Fig.\ref{rho} (a)), suggesting that it is the 3D Mott Variable Range Hopping model,
\begin{equation}
	\rho=\rho_{0}\cdot exp\bigg[\bigg({\frac{T_{0}}{T}}\bigg)^{1/4}\bigg]
\end{equation}  
that likely governs the conduction mechanism in this temperature range. This mechanism is known to describe the low temperature conduction in disordered systems with localized charge carrier states and conduction in other 6H perovskites is also reported to be behave similarly \cite{Phatak2013,Naveen2018,Kumar2021}. The 3-dimensional conduction suggests that the interlayer and intralayer hopping integrals are of similar energies and all exchange pathways participate in conduction. Below $T_f$, the material becomes highly insulating due to increased localization/freezing of these hopping electrons. 
Dielectric spectroscopy with an excitation field of $V_{ac}=1V$ in the frequency range 0.3KHz - 1MHz reveals Ba$_2$CoRuO$_6$ to be an extremely lossy dielectric for $T>150K$. Below 150K, $\epsilon'$ and $\epsilon''$ decrease smoothly and do not show any marked features, however, on close inspection, a very subtle frequency dependent feature can be seen near 50K in the loss tangent (tan$\delta=\epsilon'' / \epsilon'$) (Fig.\ref{rho} (b)). This could point to the existence of a weak magnetodielectric coupling, however, extensive magnetic field dependent dielectric measurements would be required to confirm this. 

\begin{figure*}[ht]
	\includegraphics[trim=1cm 1cm 1cm 0,width=\textwidth]{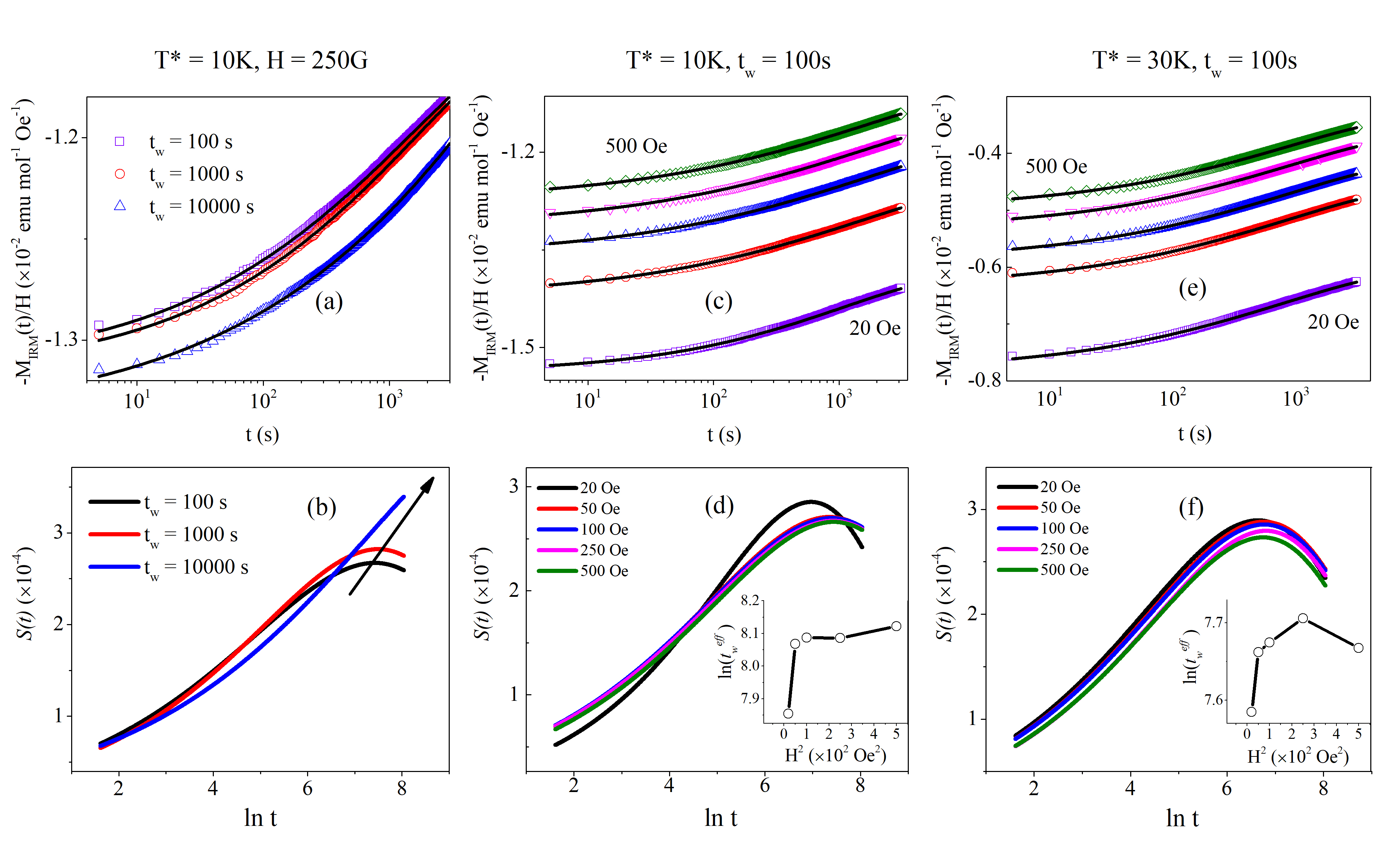}
	\caption{Time dependence of isothermal remnant magnetization (M$_{IRM}$) plotted in the form of -M$_{IRM}(t)/H$ vs time $ t $ for (a) different waiting times at 10K, and different fields at (c) 10K and (e) 30K. Solid lines denote analytical fit using a modified stretched exponential. (b), (d) and (f) show $ S(t) $ vs t calculated from the fits shown in (a), (c) and (e) respectively. Insets of (d) and (f) show a plot of $ ln(t_w^{eff}) $ vs $ H^2 $ showing the lack of variation in $t_w^{eff}$ with applied magnetic field.}
	\label{agingtw} 
\end{figure*}

To further probe the non-equilibrium relaxation dynamics of the frozen state we perform field ($H$) and waiting time ($t_w$) dependent aging experiments using the following protocol: the sample is cooled in presence of a finite field from $T>T_f$ to $T^*<T_{f}$ and after waiting for time $t_w$, the field is switched off and the magnetization is allowed to relax. We measure this relaxation with time for $\sim2$ hours. This isothermal remnant magnetization ($M_{IRM}$) follows the stretched exponential function
\begin{equation}
	M_{IRM}(t) = M_0 + A \cdot exp\big( - \big( \frac{t}{\tau}\big)^{1-n}\big)
\end{equation}
which is typically used to describe the slow dynamics of spin glasses. Here, $M_0$ is the static magnetization, A is the relaxing, glassy component, $\tau$ is the characteristic relaxation time and $n$ is the time-stretching exponent. The value of $n$ critically governs the exact relaxation rate and is expected to lie around 0.5 for spin glasses. For all our fits for different $t_w$ and $H$ (Fig.\ref{agingtw} (a),(c) and (e)), the value of $n$ is always found to vary between 0.5 - 0.6, confirming that the relaxations are slower than exponential. At a given temperature, $t_{w}$ is proportional to the size of the frozen domains. Waiting for a longer time allows one to probe the aging behavior of larger domain sizes \cite{suzuki2008} with longer relaxation times. As a result, when the waiting times are increased, so do the effective response times of the large clusters \cite{Lundgren1983}. The response function $S(t)$ given by:
\begin{equation}
	S(t) = \frac{d}{d ln(t)}\bigg(\frac{-M_{IRM}(t,t_w)}{H}\bigg)
\end{equation}
is expected to peak at a time of the order of $t_w$. This is reflected in Fig.\ref{agingtw}(b) where $S(t)$ peaks can be seen shifting towards longer timescales with increasing waiting times. The only exception is found for $t_w=100$ s where $S(t)$ peaks around the same time as $t_w=1000$ s albeit slightly earlier. This could be related to the time required for temperature and field stabilization during the measurement, which would affect the lower waiting times more, since these two timescales are of similar orders of magnitude. 

We also probe the magnetic field dependence of aging to estimate the SG correlation length from the variation of the effective waiting time $t_w^{eff}$ (the time at which $S(t)$ peaks) with $H^2$. $S(t)$ is linked to the overall height of the energy barriers available which changes as a function of the external magnetic field. When the field is increased, these effective barrier heights are reduced and thus, smaller values of $t_w^{eff}$ are expected as the field is increased \cite{Joh1999,Vincent1995}. Curiously, we see $S(t)$ peaking at (slightly) higher timescales with increasing $H$ (Fig.\ref{agingtw}(d),(f)). However, a closer look reveals that the change is miniscule (of only a few tens of seconds). This means $ S(t) $ does not appreciably shift its peak from the vicinity of $t_w$ between 20-500 Oe and the value of $\tau$ essentially remains unchanged ($\sim10^3s$) through different external fields. This is interesting since the magnetic (Zeeman) energy ($E_z$) associated with a change in field $H$, is related to the volume over which the spins are locked together for barrier hopping ($N_s$) as $E_z = N_s \chi_{FC}H^2$ where $\chi_{FC}$ is the field cooled susceptibility per spin. The spin glass correlation length $\xi$ is the radius of this volume $N_s$. If $N_s$ is built of smaller clusters, the activation energy corresponds to smaller barrier heights and $ S(t) $ always peaks around the waiting time $t_w$ (which is fixed), thus returning similar $t_w^{eff}$ for different applied fields \cite{Joh1999}. This is precisely what we find in our measurements and this validates our claim of a cluster-glass ground state for this system. However, since no field dependence was found, we were unable to calculate the SG correlation length using this method.

\begin{figure}[ht]
	\includegraphics[trim=1cm 1cm 0 1cm, width=\columnwidth]{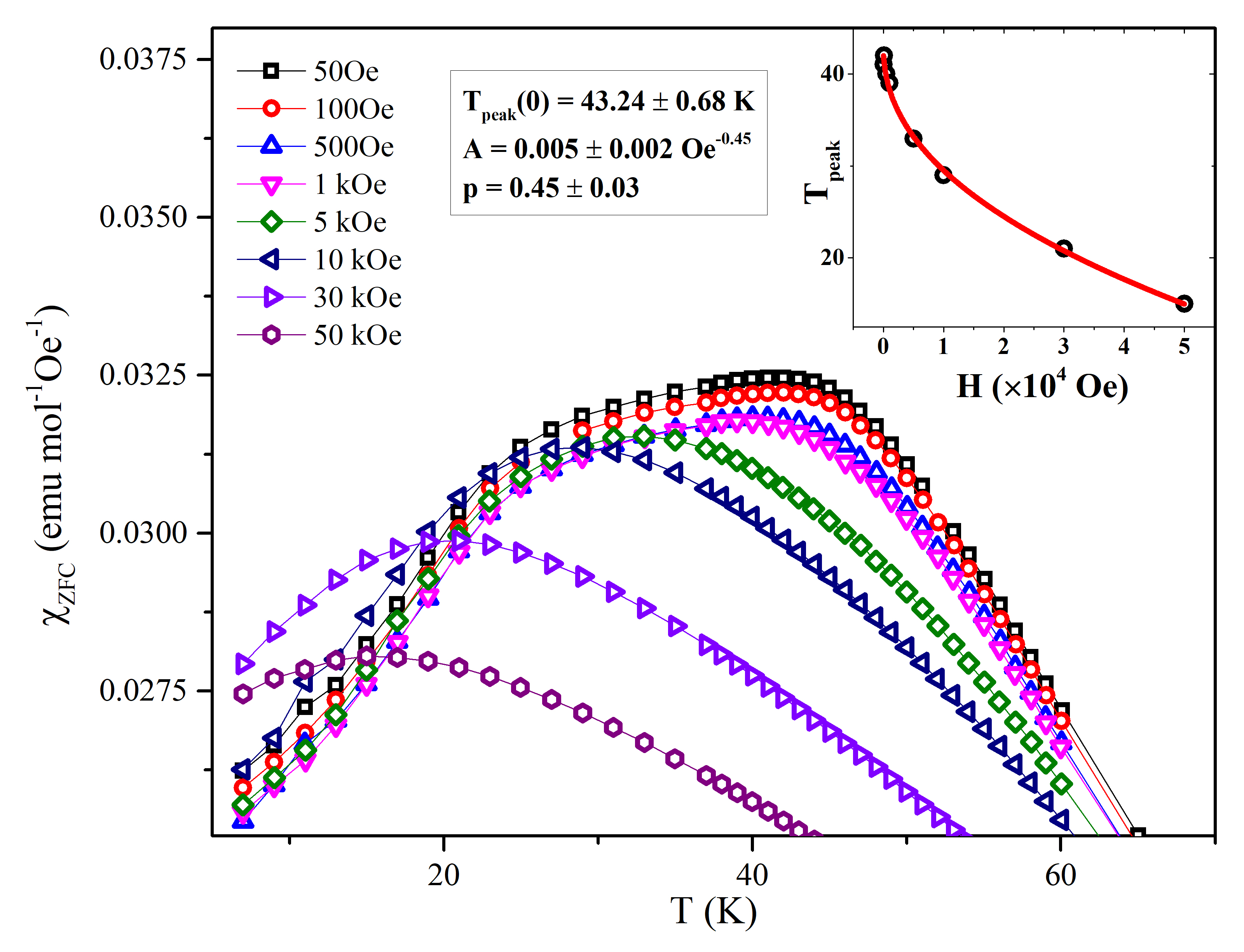}
	\caption{Suppression of the ZFC maxima with increasing magnetic field $ H$ = 50 Oe to 50kOe. Inset: Temperature corresponding to the ZFC peak ($T_{peak}$) plotted against applied field $H$ and fitted to the equation $ T_{f}(H) = T_{f}(0)(1-AH^{p}) $, parameters obtained from the fit are displayed alongside.}
	\label{ZFCH} 
\end{figure}

Magnetization measured under various applied fields reveals that $T_{f}$ is strongly suppressed at higher fields (Fig.\ref{ZFCH}). Fitting $T_{f}(H)$ (Fig.\ref{ZFCH}) to the mean-field relation for vector spin glasses with random anisotropy \cite{Kotliar1984}:
\begin{equation}
	T_{f}(H) = T_{f}(0)(1-AH^{p})
\end{equation}
gives $p\approx 0.45$ and T$_{f}(0)\approx42.8K$. $T_{f}(0)$ is the temperature where the ZFC curve is expected to peak in absence of an external magnetic field, A is a constant parameterizing the randomly distributed exchange term and $ p $ is an exponent that depends on the strength of anisotropy with respect to the magnetic field. When anisotropy is strong, $p=\frac{2}{3}\approx0.67$, corresponding to the A-T (de Almeida \& Thouless) line for Ising spins \cite{AT1978}, is expected. The G-T (Gabay \& Toulouse) line \cite{Gabay1981} with $p\approx2$ is expected in Heisenberg spin systems with weak anisotropy where two transitions are predicted to mark the distinct freezing of transverse and longitudinal spins. Our value of $p\sim 0.45$ is closer to the A-T value suggesting that we are in the strong anisotropy regime. This anisotropy is most likely a result of the globally disordered distribution of the various exchange interactions present in this system, forming short-range ordered regions of varying sizes, which continue to compete with each other down to the lowest temperatures, neither being able to establish a long-range ordered ground state.  

\begin{figure}[ht]
	\includegraphics[width=\columnwidth]{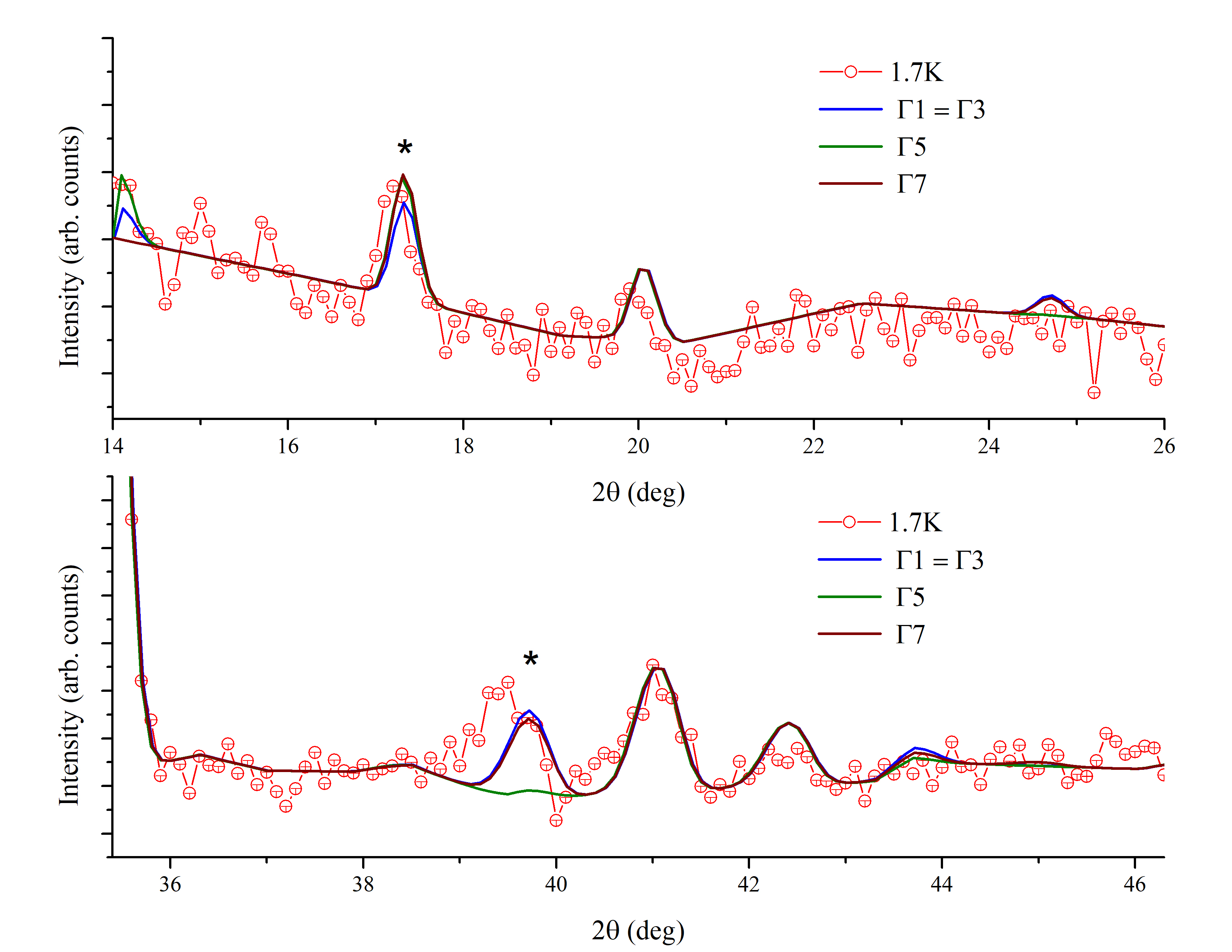}
	\caption{Neutron powder diffraction profile measured at 1.7K, showing additional peaks prohibited by the crystallographic space group $P6_{3}/mmc$ at $2\theta$ = 17.5\degree{} and 39.5\degree marked with a * and their fits to three different irreps $\Gamma1$, $\Gamma5$ and $\Gamma7$.}
	\label{LTneutron} 
\end{figure}

Neutron Powder Diffraction (NPD) patterns were also measured at temperatures 150K, 20K and 1.7K to probe the magnetic structure within the frozen spin clusters below $T_{f}$. Below 20K, two extremely weak reflections are observed at $2\theta = 17.5\degree$ and 39.5\degree. A comparison between the diffraction profile at 310K and 1.7K is shown in Fig.\ref{LTneutron} illustrating the magnetic Bragg reflections. Using the K-SEARCH program in the FULLPROF Suite \cite{fullprof}, we find that the k-vector that generates these magnetic peaks is ($\frac{1}{2}$,0,0) (equivalent to (0,$\frac{1}{2}$,0) and ($\frac{1}{2}$,$\frac{1}{2}$,0)) suggesting that an antiferromagnetic order with a magnetic unit cell doubling exists within the clusters. No significant enhancement of the crystallographically allowed peaks is observed, ruling out ferromagnetic order. However, we do observe hysteresis loops in $M(H)$ measurements which widen on cooling the sample (Fig.\ref{MT}). Although the coercivity at 2K is 1.9 kOe, no saturation is observed. This reveals the persistence of a small weak FM contribution that continues to compete with the dominant AFM interactions in the CG phase.

\begin{figure}[ht]
	\includegraphics[width=\columnwidth]{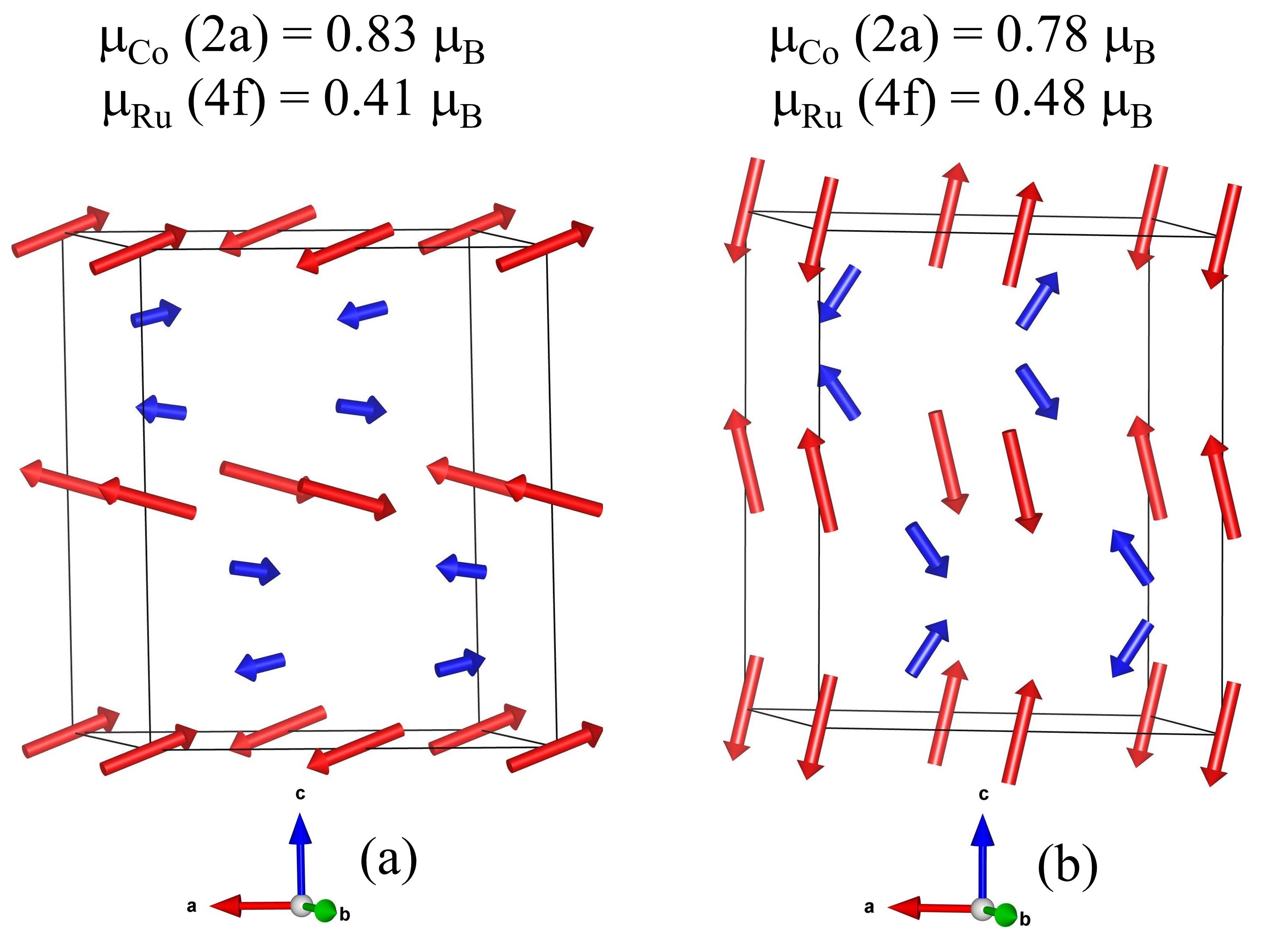}
	\caption{The two plausible spin configurations within the frozen clusters obtained from magnetic structure refinement using the irreps (a) $\Gamma_1$ and (b) $\Gamma_7$ which gave the best fits to the neutron diffraction data collected at 1.7K. Both configurations result in similar goodness of fits and moment values on the two magnetic sites.}
	\label{magstruc} 
\end{figure}

Magnetic representation analysis was performed using SARAh \cite{wills2000,wills2015} which resulted in only the odd irreps surviving for the corner sites (6 basis vectors were projected), while all the irreps survived for the dimer sites (12 basis vectors were projected). On testing with various combinations of these irreps for both the sites, we eventually could see that $\Gamma_1$ and $\Gamma_3$ were identical to each other and both $\Gamma_1$ and $\Gamma_7$ gave good fits to both the magnetic peaks. This is shown in Fig.\ref{LTneutron}. For the sake of clarity we also show the fit using $\Gamma_5$ in Fig.\ref{LTneutron} and it can be seen that it does not generate any intensity for the peak at 39.5\degree. So, we can conclude that the fits from $\Gamma_1$ and $\Gamma_7$ are of similar quality. Fig.\ref{magstruc} shows the stuctures generated from each of these irreps. Both the structures show canted spins arranged antiferromagnetically along the c-axis and ferromagnetically along b. The major difference between the two seems to be the direction of canting. The $\Gamma_1$ structure has spins canted about the a-b plane, while in the $\Gamma_3$ structure they cant about the c-axis. Since, the moment values obtained from both these fits are similar, we cannot unambiguously identify the correct structure, and both of these are plausible. The moments within the dimers are antiferromagnetically coupled for both the structures so it makes sense that we get extremely small moments from those sites. Moment suppression is also observed on the sites solely occupied by Co and is likely the result of the strong disorder and exchange randomness immanent  within this material \cite{saha2001}.

\section{Conclusions}

We report an extensive study of the 6H disordered double perovskite Ba$_2$CoRuO$_6$ via Synchrotron and Neutron Powder Diffraction, ac and dc magnetization, waiting time and field dependent-aging, resistivity, specific heat and dielectric measurements and conclude it to be a disorder-driven cluster glass with a strong exchange anisotropy. The system shows classic glassy phase characteristics like a frequency dependence in ac magnetization, a field dependent ZFC-FC bifurcation in dc magnetization, no saturation at high fields, robust memory and aging effects as well as non-Debye behaviour in the low temperature specific heat which is otherwise featureless. Through rigorous analysis of the ac susceptibility data we establish that the participating entities here are magnetic spin clusters that start forming near room temperature and continue to grow on further cooling. At the lowest temperature of 2K, this system displays signatures of small AFM order (additional low angle peaks in neutron diffraction pattern) within the frozen clusters, along with a weak FM contribution. Within the frozen clusters, the AFM order has a doubled magnetic unit cell which is consistent with an arrangement of canted spins either about the a-b plane or the c-axis with both AFM (along a- and c-axis) and FM (along b-axis) correlations persisiting and small moment values $<1\mu_B$. This suggests that the system  harbours a fierce competition between FM and AFM interactions which is strong enough to prohibit long range ordering down to 2K. Analysis of the $T_f-H$ phase gives a $p$-value which is also consistent with a system containing strong random anisotropy, even though it does not mark the existence of the A-T line. This exchange anisotropy promoted by the large disorder establishes a very frustrated network of Co$^{3+}$ and Ru$^{5+}$ ions. On cooling, strengthened local correlations lead to clustering of these ions in random regions forming both ferro- and antiferromagnetic domains which grow in size and strength until they undergo a cooperative global freezing at $T_f \sim 43K$ and form a cluster glass ground state. \\ 

\begin{acknowledgments}
SC and SN thank the Department of Science and Technology, India (SR/NM/Z-07/2015) for financial support and the
Jawaharlal Nehru Centre for Advanced Scientific Research (JNCASR) for managing the project. SC and SN acknowledge support from an Air Force Research Laboratory grant (FA2386-21-1-4051). SC is thankful to Dr. J.R. Carvajal and Dr. Sudhindra Rayaprol for extensive discussions on magnetic structure solving using neutron diffraction data. SC and SN thank Prof. A. K. Nigam for specific heat measurements. SC is thankful to Dr. Jitender Thakur for performing synchrotron measurements at PFKEK (Japan). This work is partly based on experiments performed at the Swiss spallation neutron source SINQ, Paul Scherrer Institute, Villigen, Switzerland. \\
\\
\\
\end{acknowledgments}

\bibliography{Ba2CoRuO6}

\end{document}